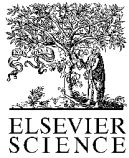
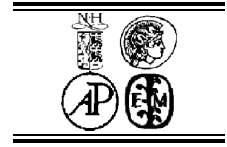

# Doping dependent time-reversal symmetric nonlinearity of YBa$_2$Cu$_3$O$_{7-\delta}$ thin films

Sheng-Chiang Lee and Steven M. Anlage[*]

*Center for Superconductivity Research,
Physics Department, University of Maryland, College Park 20742-4111
U.S.A.*

**Abstract**

We have measured the temperature dependent third harmonic response from a series of under-doped YBa$_2$Cu$_3$O$_{7-\delta}$ thin films to address the mechanism of nonlinearity in these samples. We find that the intrinsic Ginzburg-Landau nonlinearity near T$_c$ is doping dependent, with the sample becoming more nonlinear as it is under-doped. The results are consistent with the doping dependence of the condensation energy of YBa$_2$Cu$_3$O$_{7-\delta}$. © 2001 Elsevier Science. All rights reserved

*Keywords: microwave; nonlinear; superconductor; doping*

## 1. INTRODUCTION

The well-known Ginzburg-Landau description of superconductivity predicts an intrinsic third order nonlinearity of the superfluid response, which is enhanced tremendously near T$_c$. Another well-known third order nonlinearity of interest in high-temperature superconductors (HTSC) occurs in the Andreev Bound State (ABS). To explore and provide an alternative means to locally study these nonlinearities, we employ our near-field microwave microscope to study the local harmonic response from HTSC thin films.

The nonlinear dynamics of superconductors is manifested in different physical quantities in different ways. In the presence of a time-reversal symmetric (TRS), third order nonlinearity, the superfluid density $\rho_s$ becomes dependent on the externally applied current density $J$ through, $\rho_s(T,J)/\rho_s(T,0) \cong 1-(J/J_{NL}(T))^2$, where $J_{NL}(T)$ is the temperature-dependent, but experimental system-independent, scaling current density, which gives a quantitative measure of the nonlinear mechanism. Each mechanism of nonlinearity predicts its own value and temperature dependence for this scaling current density. For example, in the Ginzburg-Landau (GL) description of s-wave superconductors, $J_{NL}(t)$ is expected to be $J_{c0}(1-t^2)(1-t^4)^{1/2}$, where $J_{c0}$ is on the order of $10^{13}$ A/m$^2$ in HTSC, and $t = T/T_c$ is the normalized temperature.

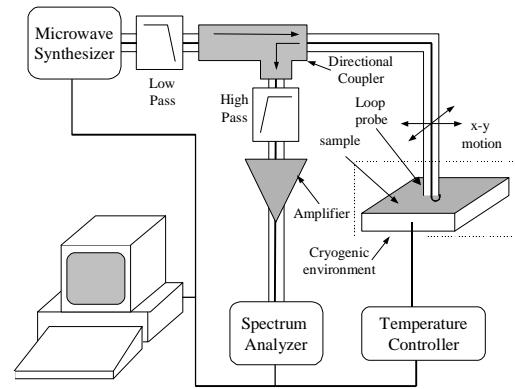

Fig. 1 System schematic. See Ref. [1] for details.

## 2. RESULTS

Details of the experimental setup (Fig. 1) are described elsewhere [1]. In summary, we apply microwave signals locally to the sample through a loop probe to induce localized microwave currents on the surface, and measure the third harmonic power ($P_{3f}$) in the signals reflected back from the sample.

---

[*] Steven M. Anlage. Tel.: 1-301-405-7321; fax: 1-301-405-3779; e-mail: anlage@squid.umd.edu.

The samples under study are a series of under-doped YBa$_2$Cu$_3$O$_{7-\delta}$ (YBCO) thin films, with thickness around 1000Å. The hole-doping concentration x is estimated from the $T_c$ of each film, using the formula $T_c = 92K(1 - 82.6(x - 0.16)^2)$. The films were originally deposited on NdGaO$_3$ or SrTiO$_3$ substrates by the pulsed laser deposition technique at nearly optimal doping. Some of the films were subsequently re-annealed at different temperatures and oxygen pressures to change the oxygen content. All films have a transition width less than 5 K, as determined by ac susceptibility measurements.

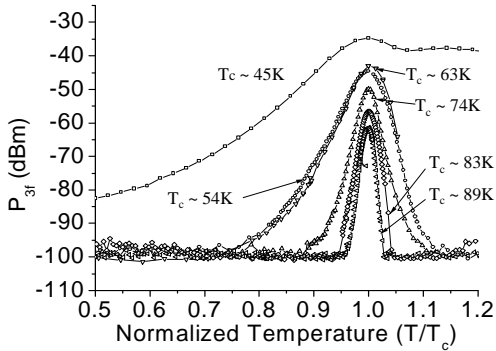

Fig. 2 $P_{3f}(t)$ of under-doped YBCO films versus normalized temperature.

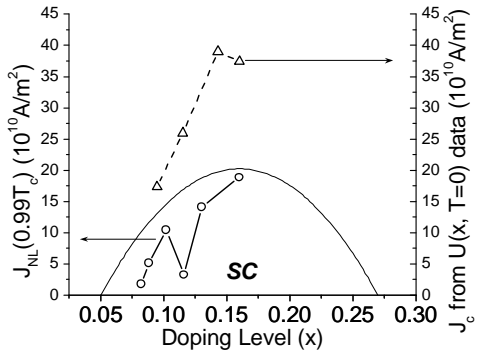

Fig. 3 Extracted $J_{NL}$ at $0.99T_c$ of under doped YBCO films versus doping level (circles). Also shown is $J_c$ deduced from condensation energy data $U(x)$ at $T = 0$ K (triangles).

In Fig. 2, measurements of the temperature dependent $P_{3f}$ of a series of under-doped YBCO films are presented versus normalized temperature. The $T_c$ values are determined by the peak $P_{3f}$ response temperature. The peaked behavior of $P_{3f}(T/T_c)$ near $T_c$ is generally expected by all models of superconducting nonlinearity. Note the trend that $P_{3f}(T/T_c)$ gets stronger and broader around $T_c$ as the samples are more under-doped (with one exception in Fig. 2). The strong persistence of $P_{3f}$ above $T_c$ in the under-doped materials is consistent with the observation of $\sigma_2 > 0$ above $T_c$ in other under-doped materials [2].

## 3. DISCUSSION

Using the algorithm described elsewhere [1], we can extract the $J_{NL}(T/T_c)$ which gives a model- and system-independent means of quantifying the evolution of nonlinearity with temperature and doping. In Fig. 3, we present the extracted $J_{NL}(0.99T_c)$ at different doping levels from the $P_{3f}$ data presented in Fig. 2, employing a doping-dependent penetration depth deduced from the literature [3].

Fig. 3 clearly shows a trend that the $J_{NL}(x, 0.99 T_c)$ decreases with decreasing x. Here we consider an interpretation of $J_{NL}(x)$ in the context of GL theory, where it is the depairing critical current density $J_c(T)$. This critical current density should be on the scale of $10^{13}$ A/m$^2$ at an intermediate temperature range, consistent with our results. On the other hand, this quantity also relates to the condensation energy U of the superconductor through $J_c \sim H_c/\lambda$, and $U \propto H_c^2$, where $H_c$ is the thermodynamic critical field, and $\lambda$ is the GL magnetic penetration depth. According to the work of Luo *et al.* [4], $U(x)$ is peaked near optimal doping and decreases as the sample becomes under- or over-doped. Taking the doping-dependent $\lambda$ into consideration, we find the doping dependence of $J_{NL}$ determined from our third order nonlinearity measurements is consistent with the $J_c(x, T = 0)$ deduced from the condensation energy results (Fig. 3).

We have measured the third harmonic response from HTSC thin films to study the local nonlinear mechanisms. We studied the trend of $J_{NL}$ with doping level of the sample, and find that the samples are more nonlinear as they are under-doped. Our results are consistent with the doping dependent condensation energy U of YBCO.

This work was supported by NSF-GOALI DMR-0201261 the Maryland/Rutgers NSF MRSEC under DMR-00-80008, and NSF IMR under DMR-98-02756. We also thank Matt Sullivan for making the films, Ben Palmer for oxygen annealing and Greg Ruchti for supporting numerical calculations.

## 4. REFERENCES

[1] S.-C. Lee and S. M. Anlage, Appl. Phys. Lett. **82** (2003) 1893.
[2] J. R. Waldram, D. M. Broun, D. C. Morgan, R. Ormeno, and A. Porch, Phys. Rev. B **59** (1999) 1528.
[3] C. Panagopoulos, J. R. Cooper, T. Xiang, Y.S. Wang, and C. W. Chu, Phys. Rev. **B** 61 (2000) 3808.
[4] J. L. Luo, J. W. Loram, J. R. Cooper and J. Tallon, Physica C341-348 (2000) 1837.